\documentclass[journal, 12pt, onecolumn]{IEEEtran}
\IEEEoverridecommandlockouts
\usepackage{graphicx}
\usepackage{listings}
\usepackage{wrapfig}
\begin{document}

\title{Leveraging MPI RMA to optimise halo-swapping communications in MONC on Cray machines}

\author{
    \IEEEauthorblockN{Nick Brown, Michael Bareford, Mich\`{e}le Weiland}
    \IEEEauthorblockA{EPCC, University of Edinburgh, Edinburgh, United Kingdom
    \\n.brown@epcc.ed.ac.uk, m.bareford@epcc.ed.ac.uk}
}

\maketitle

\begin{abstract}
Remote Memory Access (RMA), also known as single sided communications, provides a way of accessing the memory of other processes without having to issue explicit message passing style communication calls. Previous studies have concluded that MPI RMA can provide increased performance over traditional MPI Point to Point (P2P) but these are based on synthetic benchmarks. In this work, we replace the existing non-blocking P2P communication calls in the MONC atmospheric model with MPI RMA. We describe our approach in detail and discuss options taken for correctness and performance. Experiments on illustrate that by using RMA we can obtain between a 5\% and 10\% reduction in communication time at each timestep on up to 32768 cores, which over the entirety of a run (of many timesteps) results in a significant improvement in performance compared to P2P. However, RMA is not a silver bullet and there are challenges when integrating RMA  into existing codes: important optimisations are necessary to achieve good performance and library support is not universally mature. In this paper we discuss, in the context of a real world code, the lessons learned converting P2P to RMA, explore performance and scaling challenges, and contrast alternative RMA synchronisation approaches in detail.
\end{abstract}


\section{Introduction}
The Met Office NERC Cloud model (MONC) \cite{easc} is an open source high resolution modelling framework that employs large eddy simulations to study the physics of turbulent flows and further develop and test physical parameterisations and assumptions used in numerical weather and climate prediction. Written in Fortran 2003, MONC replaces an existing model called the Large Eddy Model (LEM) \cite{lem} which was an fundamental tool, used by the weather and climate communities, since the 1980s for activities such as development and testing of the Met Office Unified Model (UM) boundary layer scheme \cite{lock1998}\cite{lock2000}, convection scheme \cite{petch2001}\cite{petch2006} and cloud microphysics \cite{abel2007}\cite{hill2014}. 

The current MONC model has been demonstrated up to 32,768 cores \cite{easc} but the intention over the next couple of years is to scale this up to over a hundred thousand cores which will enable scientists to tackle some of the grand challenges in atmospheric science. However, going to this much greater scale will undoubtedly push (and possibly exceed) the parallelisation and communications technologies currently utilised by MONC, namely MPI Point to Point (P2P). A major question therefore is whether other communication technologies can be leveraged that provide greater scalability.

In this paper, we describe work done replacing the existing MPI P2P communications with Remote Memory Access (RMA) for halo-swapping in MONC. RMA is often closer to the physical hardware representation of communication \cite{PerformanceMPIRMA}, which means RMA data can often be put directly \emph{on the wire}, whereas P2P requires layers of abstraction on-top of the hardware which introducing overhead. Additionally MPI P2P messages often require overhead in the library, especially on the receiver side, where matching must be performed. For instance the tag, which can be wild-carded, along with communicator must be carefully checked to ensure a message is being delivered correctly. As RMA is one-sided, and the communications themselves far simpler, then much of the library messaging overhead can be avoided.

Whilst there have been numerous benchmark studies focusing on the benefits of MPI RMA over P2P \cite{PerformanceMPIRMA}\cite{RMAperformance}\cite{RMAimplperf}\cite{rmacug}, there are a lack of real world applications that have been converted to use MPI RMA. In this paper, driven by our work on the MONC code, we consider both the performance benefits of MPI RMA over P2P and also the approach we have adopted which helps further enhance community best practice in this area.

The rest of this paper is organised as follows; section 2 discusses the context of MONC and gives an overview of MPI RMA. Following this, section 3 provides a short description of the current halo-swapping mechanism in MONC and specifically the API that other parts of the model interact with. Section 4 focuses on our replacement of MPI P2P communication with MPI RMA, here we discuss the approach taken to leverage all three synchronisation modes of MPI RMA whilst ensure correctness and targeting performance. Weak and strong scaling experiments have been run and results of our RMA implementation using a standard MONC test case on up to 32,768 cores of a Cray XC30 are compared against the existing P2P code in section 5. In section 6 we detail reruns of these experiments on an SGI ICE machine to enable us to understand whether the observations on the Cray are universal or machine specific. Finally, we present our conclusions and discuss further work in Section 7.

\section{Background}
\label{sec:bg}
MONC has been shown to scale well up to 16384 cores \cite{easc} but as we reach 32768 cores, the overhead of communication starts to become more significant. The model works on a number of prognostic (raw data) fields and common configurations contain between 20 and 50 fields. The grid is distributed in two dimensions across processes and each process holds every field for its own specific part of this grid. Halo-swapping is the major form of communication, with a box stencil (which includes the corners) of depth two. Hence for all halo-swaps each process must communicate with up to eight neighbours (left, right, up, down and all four corners.) Halo-swapping occurs at multiple points during a timestep:

\begin{itemize}
	\item At the start of each timestep all fields must be halo-swapped before the other activities of the timestep can proceed. This means that the ability to overlap communications and compute is fairly limited here as the halo values must be present for the rest of the computation. Indeed this aspect accounts for around 95\% of the time spent doing communication per timestep.
    \item TVD advection \cite{tvd} is the major way in which values are moved through the atmosphere due to wind. This requires a halo-swap in order to process the last column of the domain for all fields. For this we can overlap communications and compute as we just need to halo-swap in one direction by sending appropriate values to the proceeding neighbour (going left in the domain.) Once a process has calculated the first column, it send the values to its proceeding neighbour and registers a receive for the halo at the other end of the domain from the succeeding neighbour. Whilst communications are on-going calculations are performed for the middle of the domain and we just need to wait for the halo values when we get to the last column. Hopefully by that point the communication will have completed and the code can directly progress onto processing these.
    \item The pressure calculation requires multiple halo-swaps. Firstly, source terms are calculated and a halo-swap is required at this point. Once source terms are computed the calculation of pressure terms involves solving the Poisson equation for diffusion via an iterative solver \cite{poisson-iterative}. This iterative solver requires a halo-swap for each iteration.
\end{itemize}

MONC has a modular design where the majority of functionality is contained in a suite of independent components. These components \emph{plug-in} to the model, driven by the user's configuration, and as the model progresses can be called at different points (for instance during model initialisation, for each timestep and model finalisation.) Components provide the scientific functionality and high level parallelism with a separate \emph{model core} providing centralised component management, such as component registration and marshalling, as well as utility functionality to avoid duplication of common code in components. Example of this utility functionality includes logging, data format conversions and maths operations. 

As described in this section, several components require halo-swapping. So in order to avoid code duplication the model core implements the \emph{mechanism} of halo-swapping and components call into this providing their own policy (i.e. which fields to halo-swap.) The idea is that the complexity involved with halo-swapping is hidden by the mechanism utility code of the model core, which contains the actual MPI calls, and components can simply call into this without the scientific programmer needing a deep understanding of the underlying nature of communications.

Hence multiple components in MONC provide the \emph{policy} of halo-swapping, directing what fields to swap, and the model core contains the underlying \emph{mechanism} that actually implements the communication in an efficient manner.

A major benefit of this approach to our work is that, in terms of replacing the MPI P2P with RMA for halo-swapping, there is a single point of truth - utility functionality provided by the model core. By modifying this one central location then all parts of the model that perform halo-swapping will be able to take advantage of the changes. However a challenge is that the API (the interface to halo-swapping mechanism) has been designed with P2P communications strongly in mind. Hence the way in which components provide their policy to the halo-swapping mechanism in the model core works well for P2P communications, but we don't want to modify this API because it would require considerable changes elsewhere in the model. Instead we want to keep the API, namely the interface the mechanism presents to the rest of the world, exactly the same so that components can transparently take advantage of MPI RMA.

\subsection{MPI Remote Memory Access (RMA)}
MPI Remote Memory Access (RMA) is a way of reading and writing directly to the memory of other processes without having to go through the point to point semantics of communication. Commonly known as one-sided communication, memory is exposed between processes via \emph{windows} and a collective call creates a window on each process in the provided communicator. These windows of memory are then used as a basis for one-sided RMA communications. 

All communication operations in MPI RMA are non-blocking and these are issued inside \emph{epochs}. Stopping an epoch will block for all communication operations issued within it to complete. These communications (most commonly \emph{put} to write data to remote memory and \emph{get} to read from remote memory) are issued from an \emph{origin} process and interact with the memory of a \emph{target} process. In MPI RMA epochs drive the synchronisation of processes and there are three approaches to synchronisation; \emph{fence}, \emph{Post-Start-Complete-Wait (PSCW)} and \emph{passive target synchronisation (locks)}.

\subsubsection{Fence}
A fence is the simplest form of RMA synchronisation and most closely follows a barrier approach. When calling \emph{MPI\_Win\_fence} each process will synchronise with every other process in the window's communicator and the call stops the current epoch and starts a new one. The fence call supports optional assertions, where the programmer may provide hints to the MPI library about the nature of the epoch and communications for optimisation purposes, although MPI is free to ignore these. For instance the \emph{MPI\_MODE\_NOPUT} assertion informs MPI that the window will not be updated by any remote write operations in the epoch that is being started.

Whilst a fence is the simplest approach to RMA, the fact that processes must explicitly synchronise with every other process in the window's communicator means that there is often an overhead associated with this style of synchronisation.

\subsubsection{Post-Start-Complete-Wait (PSCW)}
\label{sec:pscw-bg}
In the fence approach each epoch is the same, however for optimisation MPI actually provides two types of epoch; an \emph{access epoch} and an \emph{exposure epoch}. An access epoch is used to access the remote memory of another process and RMA operations (such as \emph{put} and \emph{get}) can only be issued within an access epoch. An exposure epoch exposes memory to another process so that this remote process can then interact with the memory via RMA operations. This distinction is hidden from the programmer when using fences, as a fence starts both an access and exposure epoch. 

When using PSCW the programmer must be explicitly aware of the difference between these two types of epoch and they also have explicit control over the processes involved in the epoch, whereas with a fence it is global across the window's communicator. The \emph{MPI\_Win\_post} call will start an exposure epoch and a \emph{MPI\_Win\_wait} then stops this exposure epoch (all remote operations on the memory will have completed.) A \emph{MPI\_Win\_start} call starts an access epoch and \emph{MPI\_Win\_complete} stops the access epoch (all issued RMA operations in this access epoch will have completed.)

The \emph{post} and \emph{wait} calls accepts a group of processes, which defines the processes involved in the epochs. It can be beneficial for many communication patterns, where a process need not communicate with all other processes, but instead a limited subset. This is the case with halo-swapping in MONC, where communications are nearest neighbour rather than across all processes and being able to limit the epoch, and hence synchronisation, to neighbouring processes is useful.

\subsubsection{Passive target synchronisation (locks)}

Both fences and PSCW assume \emph{active target} synchronisation, where the target process is explicitly involved in the synchronisation by creating an exposure epoch. In the active approach only the origin issues the RMA data transfer operations, but the target is involved the synchronisation. MPI also provides \emph{passive target} synchronisation, where only the origin process is involved in the synchronisation and there is no interaction required on the target process whatsoever. In the passive approach, effectively only access epochs are started and the exposure epochs are implicit. Passive target synchronisation can be especially useful for irregular and unpredictable communications. 

Passive target synchronisation is achieved by using MPI's \emph{MPI\_Win\_lock} to start an access epoch and \emph{MPI\_Win\_unlock} to stop the epoch. Inside the epoch (between \emph{lock} and \emph{unlock}) the programmer can issue RMA communication operations as normal, these are then completed for both the origin and target on the corresponding unlock issued by the origin. Locks have a type associated with them, \emph{exclusive} or \emph{shared}. \emph{Exclusive} means only one process at a time can hold the lock and hence access the window of the target process, whereas \emph{shared} locks enable multiple processes to hold the lock and hence any number can access the target window concurrently.

In addition to the locking and unlocking of windows there are a number of other passive target synchronisation operations supported. For instance one can acquire request handles from RMA communication operations and then wait on these without explicitly unlocking the window (stopping the epoch) and \emph{flush} all the communications currently issued in the epoch which blocks for their completion on the target. 

\subsubsection{MPI RMA memory model}
A further feature to MPI RMA is that the standard does not require cache coherence. The MPI provides the concept of public and private copies of window data. In the 
\emph{separate} memory model these two areas are distinct and effectively there are two copies of the window's data. A version is held in process's own local memory (private copy) and another, public, version that other processes interact with. Thus the data that a process can interact with is separate from the copy that other processes can see. These two versions are then made consistent with a synchronisation call which, for instance, is implicit at the end of an epoch. 

The second MPI RMA memory model, introduced in MPI version 3.0, is the \emph{unified} model, where public and private copies are identical. This relies on a cache coherent machine, which in reality is pretty much all the machines used for large scale HPC. It is possible to deduce what type of model a window follows and based on this specific synchronisations might be omitted by the programmer for optimisation with the \emph{unified} model. However the cost of this is that the code will not run correctly on the \emph{separate} model. The work we describe in this paper does not make assumptions about the memory model and hence will work equally well either with the \emph{separate} or \emph{unified} memory models.

\section{Existing MONC P2P halo-swapping API}
\label{sec:existingapi}

As described in section \ref{sec:bg}, MONC is made up of many loosely coupled components which users combine together via configuration files for specific runs. The mechanism of halo-swapping is contained within a module of the model core, this provides a single API for any component to then leverage for halo-swapping communications. It is this API and underlying implementation that we are focusing on in this work. The idea being that, by making changes to this single underlying module of the model core, the benefits of MPI RMA are then provided transparently to all halo-swapping components. There are four procedures in the halo-swapping API that can be called by components:

\begin{itemize}
\item \textbf{init\_halo\_communication} - This will initiate a specific halo-swapping context. Returned back to the caller via a Fortran derived type, that can be used time and time again for the halo-swapping of specific fields. The programmer provides, via arguments, a description of their fields to be swapped. The init procedure tends to be called, by each component requiring halo-swapping, only once when the model first starts up.
\item \textbf{initiate\_nonblocking\_halo\_swap} - Starts a halo-swap with neighbouring processes as described by the provided context. This procedure first registers non-blocking receives from the process's neighbours, then packs domain data into communication buffers by copying and sends these via non-blocking sends. Buffers are required because the data in the domain is often not contiguous, hence it is copied into a contiguous area and passed into the MPI call. This is procedure is non-blocking.
\item \textbf{complete\_nonblocking\_halo\_swap} - First waits for all the communications to complete (both the sends of local data to and receiving of halos from neighbours.) Then this procedure unpacks the received data by copying it from the receive buffers into the appropriate halo locations of the domain. The completion procedure blocks until all halo-swapping communication and buffer unpacking, as directed by the context, has completed.
\item \textbf{finalise\_halo\_communication} - Cleans up memory, specifically the communication buffers, required for halo-swapping. This tends to be called when the model completes execution.
\end{itemize}

The main policy of halo-swapping that users provide to the API are the number of fields to halo-swap and procedure pointers for packing and un-packing the halo data. The halo-swapping mechanism of the model core then calls out to these user provided procedures, with the specific communication buffers, and the user's code performs the physical data copying. The idea of splitting out the initiation of completion of communication is to be able to follow the familiar pattern of overlapping communication with computation.

\section{Replacing MPI P2P with MPI RMA}
\label{sec:replacep2pwithrma}
\subsection{Initialisation}
\label{sec:rmainit}
\begin{figure}
	\begin{center}
		\includegraphics[scale=0.65]{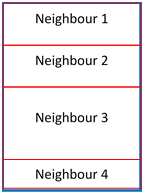}
	\end{center}
	\caption{Example buffer to be exposed via RMA}
	\label{fig:examplebuffer}
\end{figure}

Initialisation of halo-swapping is performed in the API's \emph{init\_halo\_communication} procedure. As discussed in section \ref{sec:existingapi}, the mechanism of halo-swapping relies on internal buffer space to pack non-contiguous dimensions of data for communication. The natural way of approaching this would be to create an MPI window for each buffer and then to write into these directly. However this will result in a large number of windows which would need to be managed explicitly in the code, adding to overhead, and it would not be trivial to know exactly which window a remote process should interact with. Instead, processes allocate a separate, single, buffer for RMA and create a window on this one buffer. This new buffer is large enough to hold all the data that must accessed by every neighbour and is illustrated in figure \ref{fig:examplebuffer}. It can be seen from this figure that the specific space allocated to each neighbour might not be the same, as would be the case for an uneven grid decomposition. Hence a process can not trivially know the exact location that it should work with in the remote buffer memory of a neighbouring, target, process. This is because it depends on the ordering of those neighbours at the target and the amount of remote data that the target needs to exchange with its preceding neighbours.


Therefore, during initialisation each process will calculate, for its own RMA buffer, the explicit location (offset in the buffer) that each neighbouring process should interact with. For each neighbour, a process will then exchange, via non-blocking P2P, the location in its buffer that that neighbour should interact with and it receives its own corresponding location in the neighbour's buffer (that it will need to interact with.) On the completion of this communication, each process knows not only the location in its own buffer for each neighbour, but also its own location in the remote buffer of all of its neighbours.

The \emph{MPI\_Win\_create} call is collective over the window's communicator and for this we use a communicator containing only neighbouring processes. The \emph{MPI\_Alloc\_mem} call is used to allocate the buffer space which can provide optimised memory allocation for RMA, such as ensuring memory alignment \cite{mpi-standard}.

If the user has selected to use PSCW RMA communications then an MPI group is also created that features the ranks of all the neighbours; this will be used in future calls to \texttt{MPI\_Win\_post} and \texttt{MPI\_Win\_start}.

\subsection{RMA halo-swapping}
\label{sec:rmahs}

As described in section \ref{sec:existingapi}, actual halo-swapping in the MONC API is split into two procedures; initialisation of halo-swapping, which is non-blocking, and completion which blocks for completion. With P2P halo-swapping each process sends and receives data. Sending is used to communicate values at the edge of a process's domain to the appropriate neighbours and then the corresponding halo values are received from the same neighbours. RMA is different and communication is only one way; i.e. appropriate halo data is only remotely read from neighbours (via \emph{MPI\_Get}) OR a process writes their values into the halos of remote neighbouring processes (via \emph{MPI\_Put}.) 

\begin{figure}
	\begin{center}
		\includegraphics[scale=0.55]{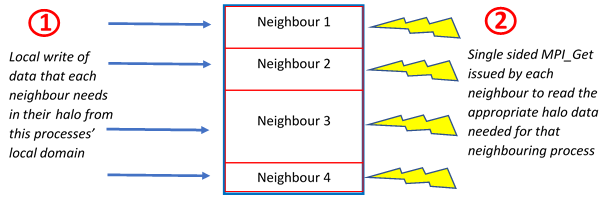}
	\end{center}
	\caption{One sided communications driven by remote reads (MPI\_Get)}
	\label{fig:getapproach}
\end{figure}

Figure \ref{fig:getapproach} illustrates the approach of remote reading from neighbouring memory to drive one-sided communications. Firstly each process must perform local copies to pack the appropriate domain data, for each neighbour, into the buffer exposed for RMA by the window. Once local copying is complete then each neighbouring process can perform a remote read (\emph{MPI\_Get}) at the appropriate location of this buffer to retrieve its corresponding data from each neighbour and place that directly into its specific halo space. For correctness the target must have completed the local copying of data (step 1) into the buffer before remote reads (step 2) are issued by other processes.

\begin{figure}
	\begin{center}
		\includegraphics[scale=0.55]{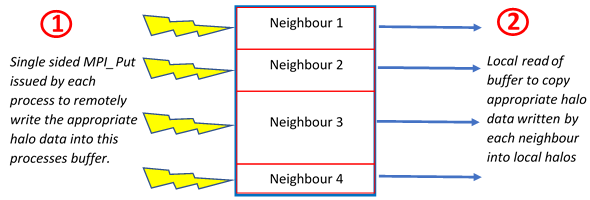}
	\end{center}
	\caption{One sided communications driven by remote write (MPI\_Put)}
	\label{fig:putapproach}
\end{figure}

The other approach is to drive the single sided communications by remote writing as depicted by figure \ref{fig:putapproach}. Here each process will remotely write (via \emph{MPI\_Put}) its appropriate data into the buffer of each neighbour at the corresponding location. Once remote writes have completed a process can read its own buffer and perform local copies to unpack the data into halos of the domain. Again, for correctness it is critical that remote writes (step 1) have completed before the buffer is read and unpacked (step 2.)

The programmer has a choice, to drive the RMA by remote reading or remote writing. But due to the one sided nature of communication, one needs to be careful about data consistency and this can be a real challenge to correctness. One needs to carefully consider the synchronisation, especially as we start to relax this in PSCW and passive target synchronisation. As we will see in this section, whether \emph{MPI\_Get} or \emph{MPI\_Put} is most appropriate can depend, in terms of data consistency, on the approach to epoch management and synchronisation that has been adopted. 

\subsubsection{Fences}

The simplest, but potentially slowest, approach for RMA synchronisation is to use fences. A fence can be placed in the \emph{initiate\_nonblocking\_halo\_swap} call to start an epoch. Then, within this same procedure, either \emph{MPI\_Get} (to remote read) or \emph{MPI\_Put} (to remote write) RMA operations can be issued and a fence can be placed at the end of the procedure to wait for their completion.

With this simple approach both of these fences are synchronisation points, starting and stopping the epoch, and both are important here for data consistency. As a fence is blocking on the window communicator, the first fence (creating the epoch) ensures that every process in the window communicator has reached this point before processes proceed. If processes then issue remote memory reads for their halo data from neighbouring processes via \emph{MPI\_Get}, we need to ensure that that data has been packed into the RMA buffer on the target before the underlying communications can begin. Placing buffer packing before the fence on each process ensures consistency as processes will only proceed once all have called the fence (similar to an MPI \emph{barrier}.)

The second fence, to stop the epoch, will ensure that all issued RMA operations inside the epoch have completed. If we choose to use remote write via \emph{MPI\_Put} instead of remote reads, then it is only after progressing beyond this point that we can be sure the RMA buffer contains the appropriate data that can be used for unpacking and copying into the local domain. 

When using fences in this manner it is tempting, for optimisation reasons, to use the \emph{MPI\_MODE\_NOPRECEDE} assertion for the first fence (as there are no locally issued RMA calls in the epoch proceeding this) and \emph{MPI\_MODE\_NOSUCCEED} as this new epoch at the end of the procedure is not issuing any RMA calls. However, crucially for correctness a call to \emph{MPI\_Win\_fence} that is known not to stop any epoch (specifically with the \emph{MPI\_MODE\_NOPRECEDE} assertion) does not necessarily act as a barrier and hence we have consistency issues if we are driving the communication via remote reads (\emph{MPI\_Get}). When modifying MONC we saw this issue intermittently in the code, as it is up to the MPI implementation whether it acknowledges the assertions. This was solved by driving communications via writing (\emph{MPI\_Put}) in place of remote reading.

For better overlapping of communication and compute, the second fence (closure of the epoch) can be placed in the \emph{complete\_nonblocking\_halo\_swap} such that the non-blocking RMA operations can then be in progress whilst other computation work is being perform between these procedures of the MONC halo-swapping API.

\subsubsection{Post-Start-Complete-Wait}
Post-Start-Complete-Wait (PSCW) can be thought of as a more efficient version of fences. Firstly processes have more control over who is involved in the epoch by the provision of groups, so instead of having to synchronise with every other process in the window communicator processes now have a more fine-grained way of determining who they have to synchronise with. Additionally this synchronisation can be further specialised by the type of epochs - an access or exposure epoch as described in section \ref{sec:pscw-bg}. 

The \emph{MPI\_Win\_post} call, used to start an exposure epoch on a window for a group of processes is non-blocking. A \emph{MPI\_Win\_start} call, used to start an access epoch on a window for a group of processes, may or may not block for a corresponding \emph{post} depending on the implementation, but it is not a requirement.This creates a challenge: if we were to drive our halo-swapping by reading remote data (\emph{MPI\_Get}) then there is no guarantee that the corresponding processes have also reached that specific point in the code and that the data in the remote RMA buffer is up to date. 

The \emph{MPI\_Win\_complete} call, which stops an access epoch, will block on locally issued RMA operations. Therefore after this call we can reuse any RMA operation buffer space on the origin, but it does not guarantee completion on the target. The \emph{MPI\_Win\_wait} call, which stops the exposure epoch on the process, will block until all the corresponding access epochs have been stopped (via \emph{MPI\_Win\_complete}) and all data from these has arrived. Hence after progressing beyond \emph{MPI\_Win\_wait} the process can be sure that it has received all the data remotely written to it in that epoch.

Based on the synchronisation behaviour of PSCW, remote writing of the halo data via \emph{MPI\_Put} is the only way in which we can guarantee correctness of the data without any additional synchronisation and-so this was the approach we adopted in the code. 

\subsubsection{Passive target synchronisation (locks)}
\label{sec:passive_replace}
In passive target synchronisation only the origin is involved in the synchronisation. The origin still starts an access epoch but unlike other MPI RMA approaches the target does not need to start an exposure epoch explicitly. This is a very useful pattern for numerous algorithms, however for our purposes of halo-swapping it is less beneficial due to the lack of explicit synchronisation on the target.

As previously discussed in this section, if the origin is reading remote data then it needs to ensure that the target has updated the RMA buffer sufficiently, or if the origin is writing remote data then the target needs to ensure that this writing has completed before it reads its own RMA buffer. Either way, some synchronisation (or at least handshaking) between the origin and target is required. 

Whilst passive target synchronisation is not necessarily most suited to halo-swapping, we were still keen to support it for understanding the performance that this approach can afford. To do this each process issues \emph{MPI\_Win\_lock\_all} in the \emph{init\_halo\_communication} procedure when halo-swapping is first set up. This is done only one for each halo-swapping context and acquires an access lock on all windows in the communicator (neighbouring processes) for the entire run. The \emph{MPI\_Win\_lock\_all} call acquires a shared lock on the window, where multiple processes are allowed to hold the lock on the same window. This is fine for our approach as each neighbouring process is working with distinct parts of the target buffer, exposed by the RMA window, so there is no conflict. The code also supplies the \emph{MPI\_MODE\_NOCHECK} assertion to the lock\_all MPI call for potential optimisation as there is no need to check for conflicting locks here.

Similarly to PCSW, remote writing is also used for passive target synchronisation and these calls are issued in the \emph{initiate\_nonblocking\_halo\_swap} procedure. In the \emph{complete\_nonblocking\_halo\_swap} procedure an \emph{MPI\_Win\_flush\_all} is issued which blocks until completion of all outstanding RMA operations initiated by the calling process on the window. The definition of \emph{operation completion} here is that the remote memory of the target has been updated and contains the specific values that were written. Hence we know that, once this call has returned, the data has arrived in the (public) memory of the target process. In the MONC halo-swapping finalisation procedure (\emph{finalise\_halo\_communication}) the code issues \emph{MPI\_Win\_unlock\_all} to release all locks on the window. With this approach we only have to acquire locks and start the access epochs once, when a specific halo-swapping context is set up and initialised. This can then be used as many times as we when when halo-swapping is being performed from timestep to timestep.

Synchronisation is a challenge here, specifically ensuring that the target does not start unpacking RMA buffer data into its halos before remote values have been fully written. To guard against this we use P2P communications with an empty message. Non-blocking receives are issued to all neighbouring processes and once these processes have flushed their RMA operations (and the data is guaranteed to be up to date on the target) then a non-blocking send of an empty message to that process is issued. A process, once it has flushed RMA operations and issued appropriate sends can then check the receive handles and unpack buffers as these operations complete (via \emph{MPI\_Testany}.) Crucially, to support the MPI separate memory mode, the target must issue \emph{MPI\_Win\_sync} before it starts unpacking the data to ensure that public and private copies of the window are up to date. As an optimisation we obtain, via \emph{MPI\_Win\_get\_attr} the memory model of the window and omit this explicit \emph{MPI\_Win\_sync} call if it is unified.

An alternative to these P2P calls would be to use an MPI barrier, but that is more coarse grained and will only complete once all processes in the communicator reach this point in code. Instead this P2P approach means that the target can progress unpacking buffers for those neighbours that have completed even if not all neighbouring processes have flushed their RMA operations. The disadvantage of passive target synchronisation here is that manual synchronisation does add some overhead in the code and puts the burden on the programmer to handle it correctly.

\subsection{Optimising epoch creation}

So far we have assumed that the epoch is started in \emph{initiate\_nonblocking\_halo\_swap} procedure and stopped in the \emph{complete\_nonblocking\_halo\_swap} procedure of the MONC halo-swap communication mechanism. Based on the discussions around data consistency in section \ref{sec:rmahs}, this allowed for a correct implementation but when profiling the code we found that excessive time was being spent by processes blocking in the \emph{initiate\_nonblocking\_halo\_swap} procedure. This was during the creation of the epoch and especially significant when using fences, but also present with PSCW where some processes were blocking on the \emph{MPI\_Win\_wait} call. 

To avoid excessive blocking we moved epoch creation out of the initiation procedure into \emph{complete\_nonblocking\_halo\_swap}. So the last thing that the completion procedure will do, after closing the existing epoch and copying RMA buffer data appropriately, is to start a new epoch that will then be used for the next halo-swap. 

Importantly, the epoch is open for much longer here. So processes reaching the \emph{initiate\_nonblocking\_halo\_swap} call at different times, which might very well happen as we aim to promote loosely coupled behaviour where possible, are not then sitting idle waiting for other processes to reach the same point. Instead they can issue their non-blocking RMA operations and progress as soon as that is done. 

The other benefit of moving the creation of the epoch to the completion procedure, is that it also fits in far better with the existing, P2P behaviour of the halo-swapping API. This is important because, with the existing API, it is only the completion procedure that blocks for data and-so programmers have developed components with this assumption in mind to overlap communication and computation as efficiently as possible. Hence also blocking when initiating the halo-swap will limit the effectiveness of these communication optimisation patterns. With the change described here, the epoch must also be started in the \emph{init\_halo\_communication} procedure so that the first call to \emph{initiate\_nonblocking\_halo\_swap} already has an epoch started and the last epoch started by \emph{complete\_nonblocking\_halo\_swap} must be stopped in the finalisation procedure.

\subsection{More effectively sharing a memory window}

The RMA communications we have described so far have used a buffer exposed for RMA via a single window as introduced in section \ref{sec:rmainit}. However the existing API design, which we did not want to modify, calls into user provided procedures (via Fortran procedure pointers) for the packing and unpacking of halo data. Crucially the buffers provided to these procedures for unpacking data (copying into the halos in the \emph{complete\_nonblocking\_halo\_swap} procedure) are assumed to be separate, multi-dimensional arrays, and not one single large buffer. This works well for MPI P2P as the code is receiving into separate buffers, but does not match so closely with how our RMA implementation works.

Due to the fact that these are multi-dimensional arrays, passing in sub-regions of the single buffer exposed for RMA into the unpacking procedures it resulted in significant overhead. This is because, instead of just passing a memory reference, Fortran copied the data by allocating new memory areas for the unpacking procedure, copied the argument's input data into these and then performing a copy back out on procedure completion. 

\begin{figure}
	\begin{center}
		\includegraphics[scale=0.55]{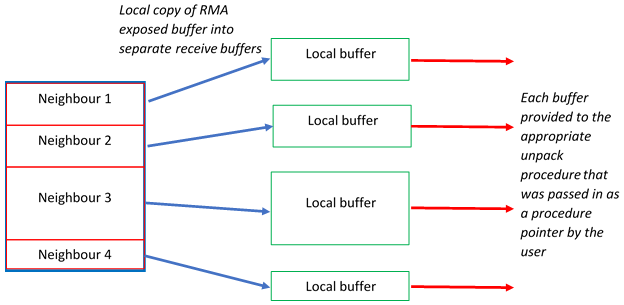}
	\end{center}
	\caption{Extra copy needed from the RMA buffer into the separate receive buffers}
	\label{fig:buffer-issue}
\end{figure}

To get round this overhead we initially copied values from the single buffer used for RMA to separate individual receive buffers. These were then passed to the user's unpacking procedures. This is illustrated in figure \ref{fig:buffer-issue}. Whilst extra data copying was required, at-least this way we could explicitly control what copying was performed, and it was quicker than passing sub-regions of the array. However the extra copy still added significant overhead, not to mention the additional memory requirements of these extra individual receive buffers.

\begin{figure}
	\begin{center}
		\includegraphics[scale=0.55]{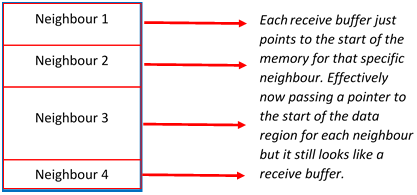}
	\end{center}
	\caption{Receive buffers now directly point into the buffer exposed for RMA to avoid extra copying}
	\label{fig:buffer-issue-fixed}
\end{figure}

Data that arrives in the single RMA buffer from a specific neighbour is contiguous and the receive buffers, provided to the user's unpacking procedures, are declared as Fortran pointers. To avoid the need for extra receive buffer space and copying, but to keep the API unchanged, the receive buffers were no longer themselves allocated. Instead they pointed to specific offsets in the single buffer used for RMA and written to directly by neighbouring processes as portrayed in figure \ref{fig:buffer-issue-fixed}. Hence each of these receive buffers is now effectively just a chunk of memory within the already allocated far larger single buffer used for RMA and the specific pieces of the single RMA window buffer are directly passed to these user provided unpacking procedures where they appear like separate chunks of memory.

The complication is that Fortran does not directly support pointer arithmetic, which is required here, so instead we must leverage C to point into the different locations of the single RMA buffer. The ISO C bindings, introduced in Fortran 2003 provide the \emph{c\_ptr} derived type which we used to keep track of the RMA memory buffer address (returned in the code from the \emph{MPI\_Alloc\_mem} call.) We wrote a simple C function, and linked this into the Fortran code, to do pointer arithmetic. This enables us to increment a \emph{c\_ptr} pointer by a specific amount for each separate receive buffer. The \emph{c\_f\_pointer} ISO C binding procedure then means we can associate the Fortran pointer of each receive buffer with the C pointer. 

\section{Cray XC30 performance results}
\label{sec:results}
Performance experiments were performed on the ARCHER, the UK national supercomputer, which is a Cray XC30 with 12-core (2.7 GHz) E5-2697 v2 (Ivy Bridge) series processors. The ARCHER compute nodes contain two Intel processors and 64 GB of memory. The MONC model was compiled using the Cray Fortran compiler v8.4.1. The MPI library used was Cray MPICH v7.5.5. Unless explicitly stated, results on ARCHER run MPI RMA over DMAPP. A standard MONC test-case for stratus cloud was used which contains 25 Q (moisture) fields, as well as temperature, pressure and wind fields. All of these need to be halo-swapped at least once per timestep. For all experiments we execute one MPI process per core. The results in this section are averaged across all timesteps of the model and multiple runs, the variance from timestep-to-timestep and run-to-run being negligible.

\begin{figure}[!ht]
\centering
\includegraphics[scale=0.65,keepaspectratio]{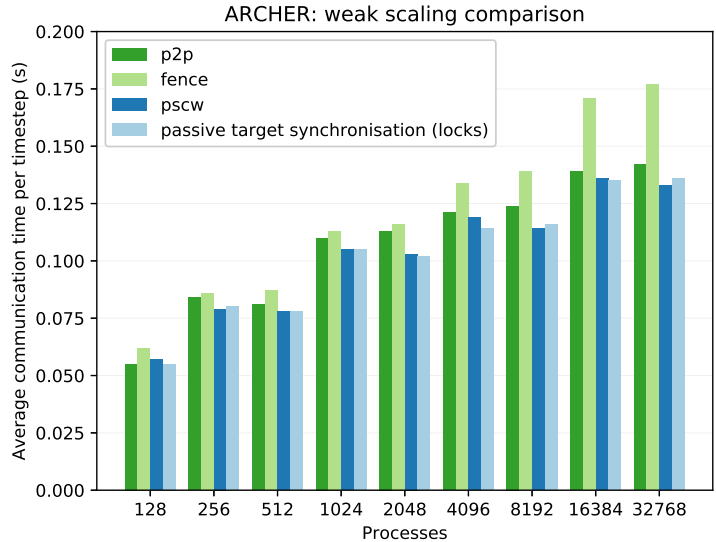}
\caption{Weak scaling (65k grid points per process) on ARCHER. Average communication time per timestep for P2P and RMA communications.}
\label{fig-archer-weak}
\end{figure}

Figure \ref{fig-archer-weak} illustrates weak scaling with a local problem size of 65 thousand grid points per process (this size is a common choice in scientific runs), for MPI P2P and all three RMA synchronisation modes. The local grid size is \emph{x=16, y=16, z=256}. The \emph{z} dimension is the vertical and the grid is not decomposed in this dimension, whereas global decomposition occurs on the other, \emph{x} and \emph{y} dimensions. Therefore processes hold a number of columns of grid points, in this case  \emph{16 $\times$ 16} (256) columns of size 256. Based on this set up, per timestep, there are \emph{16 $\times$ 256 $\times$ 2} (as the stencil depth is 2) grid points being exchanged with each left, right, up and down neighbour per field. There are also \emph{256 $\times$ 2} grid points to exchange with cornering neighbours per field. The fields themselves are double precision floating point numbers, so halo swapping faces results in message sizes of 64 KB and corners are 4 KB in size.

The metric reported is the average communication time per timestep and typical runs execute tens of thousands of timesteps, and hundreds of thousands or even millions of timesteps is not uncommon. At 32768 cores there is a global domain size of 2.1 billion grid points and over 768GB of data is being halo-swapped each timestep (approximately 23MB per process.) 

It can be seen, in all configurations apart from 128 processors, that RMA via PSCW and passive target synchronisation (locks) out performs the existing P2P communications. Even though the differences between the numbers is fairly small, this is significant over a real world run (for example going from P2P communications to either PSCW or passive target synchronisation saves over 500 seconds with 50,000 iterations on 1024 and 32768 cores.) Interesting, at smaller core counts fences are fairly competitive with the other two forms of RMA and P2P, however at larger core counts P2P out performs fences quite significantly. We believe that his is due to the extra synchronisation involved with fences and the fact that we are having to manually implement synchronisation ourselves with passive target synchronisation.

\begin{figure}
\centering
\includegraphics[scale=0.65,keepaspectratio]{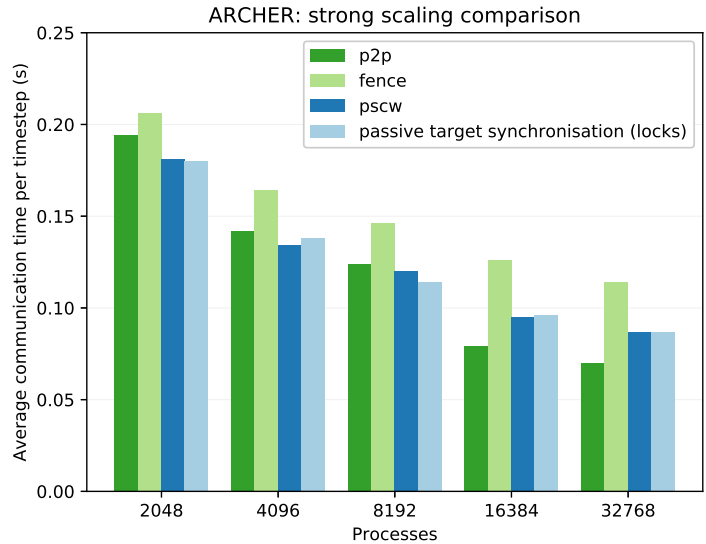}
\caption{Strong scaling (536 million global grid points) on ARCHER. Average communication time per timestep for P2P and RMA communications.}
\label{fig-archer-strong}
\end{figure}

Figure \ref{fig-archer-strong} illustrates strong scaling on ARCHER with a global problem size of 536 million grid points (global grid dimensions of \emph{x=2048, y=2048, z=128}.) This experiment is worthwhile because we can study the impact of message size on the performance of halo-swapping and figure \ref{fig-archer-strong-datasizes} shows the number of local grid points and overall amount of data communicated by a process per timestep at different core counts. Figure \ref{fig-halo-sizes-archer-strong-scaling} denotes the size of each halo that must be communicated (\emph{dimension 1} corresponds to faces communicated with left and right neighbours, \emph{dimension 2} faces to up and down neighbours and \emph{corner} to the 4 cornering neighbours) at different core counts. This is useful because it is the message size associated with the RMA operations, where the code issues \emph{MPI\_Put} operations to remote memory with data of this size. 

It can be seen that, as the local grid size becomes smaller (32 thousand points/process at 16768 cores and 16 thousand grid points/process at 32768 cores), then P2P communication becomes more competitive and is the fastest configuration at 16768 cores and very competitive with PSCW and passive target synchronisation at 32768. We found these strong scaling results noteworthy because the assumption is often that the latency of RMA is lower than P2P due to being closer to the interconnect hardware and fewer levels of abstraction. However this result seems to argue that RMA is in-fact more beneficial with larger message sizes in our case (PSCW is 8\% faster than P2P at 2048 cores, 11\% faster at 4096 cores and 5\% faster at 8192 cores.) 

\begin{figure}
\begin{center}
\begin{tabular}{ | c c c | }
\hline
 Processes & Local problem size & Data comm/timestep \\ \hline
 2048 & 262144 & 92 MB \\  
 4096 & 131072 & 46 MB \\ 
 8192 & 65536 & 23 MB \\  
 16384 & 32768 & 11.5 MB \\ 
 32768 & 16384 & 5.75 MB \\
 \hline
\end{tabular}
\end{center}
\caption{Local domain size and data communicated for each process per timestep when strong scaling (536 million global grid points) on ARCHER.}
\label{fig-archer-strong-datasizes}
\end{figure}

\begin{figure}
\begin{center}
\begin{tabular}{ | c c c c | }
\hline
 Processes & Halo size dim1 & Halo size dim2 & Corner size \\ \hline
 2048 & 128 KB & 64 KB & 2 KB\\  
 4096 & 64 KB & 64 KB & 2 KB\\ 
 8192 & 64 KB & 32 KB & 2 KB\\  
 16384 & 32 KB & 32 KB & 2 KB\\ 
 32768 & 32 KB & 16 KB & 2 KB\\
 \hline
\end{tabular}
\end{center}
\caption{RMA operation message size associated with halos when strong scaling (536 million global grid points) on ARCHER.}
\label{fig-halo-sizes-archer-strong-scaling}
\end{figure}

Whilst it is known from benchmarks that PSCW is more efficient than fences, we were still slightly surprised here by the magnitude of performance difference, especially when weak scaling over larger core counts. The reason for our surprise is that theoretically the synchronisation and epoch creation behaviour is not so different between these two RMA synchronisation mechanisms in MONC. As we have already mentioned, a fence generally synchronises with all processes in the window's communicator (in our case the neighbouring processes) although our use of \emph{MPI\_MODE\_NOPRECEDE} might avoid the first synchronisation depending on the library implementation. A fence also starts both an access and exposure epoch on each process. But our PSCW version is doing the same, again it works on just the communicator of neighbouring processes and needs to start an access and exposure epoch on each process, and the issuing of \emph{start} may or may not block, whereas the \emph{complete} and \emph{wait} calls to stop the epoch will block. So theoretically there shouldn't be too much difference, but from the results we can see that PSCW is far more efficient. One reason might be that the MPI library is ignoring the assertions provided to the fence calls, which are designed for optimisation and a conforming implementation is allowed to disregard them. Additionally the more generalised nature of PSCW might enable the MPI library to more efficiently implement the underlying synchronisations and data movements in contrast to fences.

\begin{figure}
\centering
\includegraphics[scale=0.65,keepaspectratio]{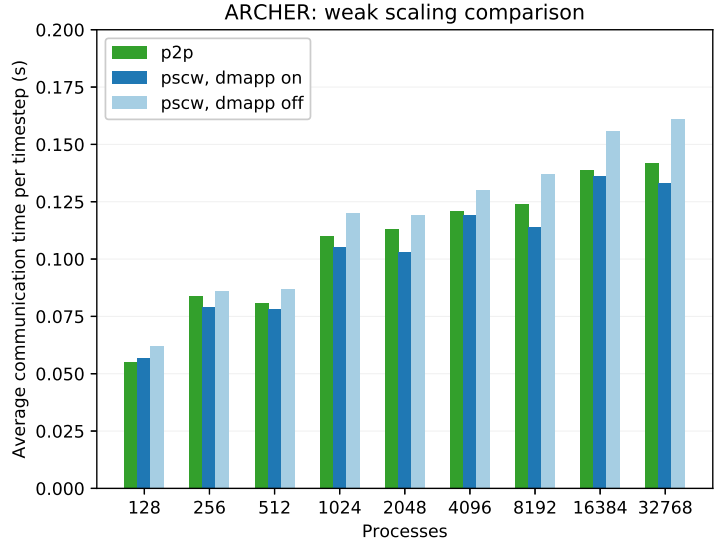}
\caption{Weak scaling (65k grid points per process) on ARCHER without DMAPP.}
\label{fig-archer-nodmapp-weak}
\end{figure}

The results reported in figures \ref{fig-archer-weak} and \ref{fig-archer-strong} rely on Cray's Distributed Memory Application API (DMAPP) \cite{dmapp} technology which is a is a communication library that can call straight through to the underlying Aries networking ASIC on the Cray and implements many of the RMA operations directly in hardware. It has previously been shown \cite{dmapp-perf} that the DMAPP interface outperforms the regular, default, MPI RMA approach. DMAPP is enabled by both linking against the DMAPP library and setting an environment variable. Figure \ref{fig-archer-nodmapp-weak} illustrates a similar weak scaling experiment as figure \ref{fig-archer-weak} but contrasting P2P communications against PSCW with and without DMAPP. It can be seen that DMAPP is certainly faster than non-DMAPP, but not by as much as we had expected and certainly not as much as benchmarks in \cite{dmapp-perf} would have led us to expect. Nevertheless, in order to obtain better performance than P2P communications using DMAPP is a must. Because DMAPP requires no code changes, it may seem like a \emph{no-brainer}, but there is also the question of fault tolerance. Due to data being transmitted directly by hardware it is more difficult to detect and fix erroneous data than via the software stack of MPI. Having said this, from trialling MPI RMA over DMAPP for a number of years, this combination now seems far more stable than previous versions were and we saw no correctness issues when running the experiments of this paper. From the results in this paper it is clear that MPI RMA over DMAPP must be enabled to obtain performance over MPI P2P communications.

\subsection{Passive target synchronisation}
The results described in this section illustrate that performance and scaling of PSCW and passive target synchronisation is fairly comparable. We thought this interesting because, as described in section \ref{sec:replacep2pwithrma}, when using passive target synchronisation the programmer must explicitly write code so that the target process doesn't read buffer data before remote writing has completed. Whereas with PSCW this is implicit in the synchronisation mode, the implementation of which is abstracted from the programmer in the underlying MPI library. From the results in this section we conclude that the performance of this synchronisation in the MPI library version is comparable with implementing it in user code.

Having said this, one still needs to be careful with passive target synchronisation. As described in section \ref{sec:passive_replace} our code calls \emph{MPI\_Win\_lock\_all} on initialisation of the model to start the epoch only once. This epoch is then stopped, again only once, when the model terminates. After communication operations for each timestep have been issued then \emph{MPI\_Win\_flush\_all} is called on the source. As also discussed in section \ref{sec:passive_replace}, the target still needs to know when its neighbours have completed remotely writing into the buffer area and hence when it can start to access halo data. For this we have used non-blocking empty P2P messages. The performance results of the implementation described here is illustrated in this section. However this was our second version of halo swapping via passive target synchronisation, with the first being simpler but also considerably slower.

\begin{figure}
\centering
\includegraphics[scale=0.65,keepaspectratio]{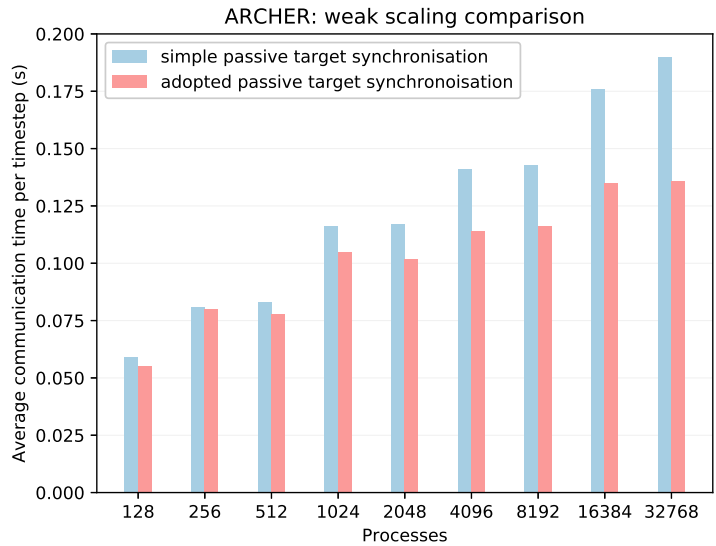}
\caption{Weak scaling (65k grid points per process) on ARCHER comparing the simple and adopted passive target synchronisation approach}
\label{fig-archer-simple-adopted-locks}
\end{figure}

Figure \ref{fig-archer-simple-adopted-locks} illustrates the performance and scaling difference between the fastest, and adopted, approach to passive target synchronisation (as described in section \ref{sec:passive_replace}) and our initial simple approach. In this first, naive, approach the epoch was started and stopped each timestep via the \emph{MPI\_Win\_lock\_all} and \emph{MPI\_Win\_unlock\_all} calls. A benefit of this was that there was no need to then flush RMA operations at each timestep, as stopping the epoch accomplishes this. Our approach to synchronisation was also much simpler, where a non-blocking MPI barrier on the neighbour communicator (so just between the neighbours) was used instead of individual empty P2P messages.

From the comparison here the reader can see, for performance, how important it is to adopt the, slightly more complicated, approach when leveraging passive target synchronisation - especially at larger process counts. The performance difference between our two approaches illustrated in figure \ref{fig-archer-simple-adopted-locks} is especially important in the context of this paper as it makes the difference between being faster or slower than an MPI P2P halo swapping approach.

\section{SGI ICE performance results}
A similar experiment to those carried out in section \ref{sec:results} was run on Cirrus, an SGI ICE machine with Intel 18-core (2.1 GHz) Intel Xeon E5-2695 (Broadwell) series processors. Each node contains two processors with 256 GB of memory and Infiniband interconnect. For this platform the MONC code was compiled with GCC v6.2.0 and the MPI implementation was SGI Message Passing Toolkit v2.14, otherwise known as MPT. Unfortunately, this MPI implementation does not support passive target synchronisation (locks) hence we can only report fence and PSCW runs for Cirrus in this section. Again for all experiments we execute one MPI process per core.

\begin{figure}
\centering
\includegraphics[scale=0.65,keepaspectratio]{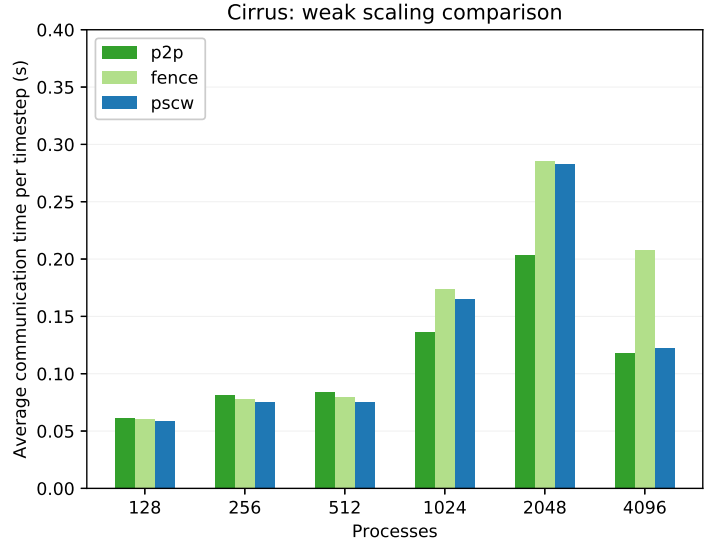}
\caption{Weak scaling (65k grid points per process) on Cirrus. Average communication time per timestep for P2P and RMA communications.}
\label{fig-cirrus-weak}
\end{figure}

Figure \ref{fig-cirrus-weak} depicts weak scaling with 65 thousand local grid points per core and the average time for halo-swapping communication per timestep. The local grid dimensions are \emph{x=16, y=16, z=256}. Whilst a direct comparison of the actual numbers between Cirrus and ARCHER is inappropriate due to different processor technology and software stack, we can make some interesting observations about the relative performance of the different configurations. Most interestingly here is that MPI RMA (PSCW) does not provide better halo-swapping performance than P2P at all core counts and at best the performance between PSCW and P2P is comparable (e.g. 512 and 4096 cores) and at worst PSCW is slower than P2P (2048 cores.)

\begin{figure}
\centering
\includegraphics[scale=0.65,keepaspectratio]{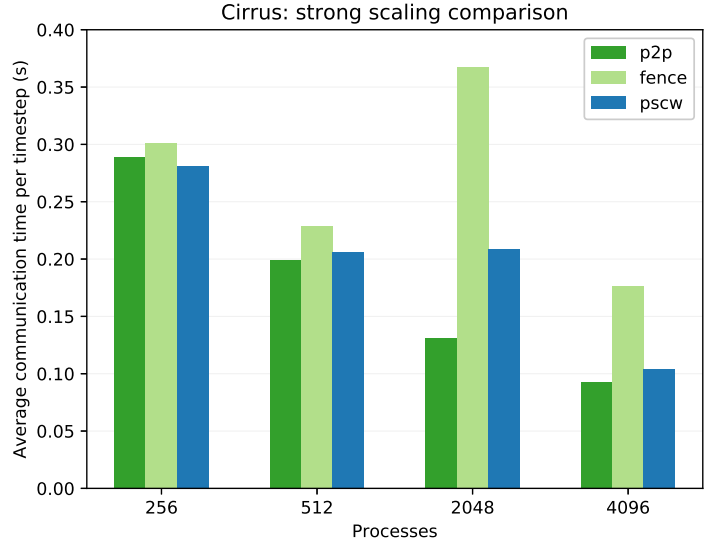}
\caption{Strong scaling (536 million global grid points) on Cirrus. Average communication time per timestep for P2P and RMA communications.}
\label{fig-cirrus-strong}
\end{figure}

Figure \ref{fig-cirrus-strong} illustrates strong scaling, 536 million global grid points, on Cirrus (global grid dimensions of \emph{x=2048, y=2048, z=128}.). Similarly to the weak scaling experiment, at best P2P and PSCW are competitive and at worst (again 2048 cores) PSCW is significantly slower than P2P. 

These results indicate that support in the SGI MPI library (MPT) and Infiniband interconnect for RMA is less mature than Cray's implementation. It is our belief that if some similar DMAPP style technology was supported on the SGI ICE machine then RMA performance would be far more competitive against P2P. However the fact that passive target synchronisation is not supported illustrates the general immaturity in MPT for RMA and-so further performance enhancements are probably unlikely in the near future.

\section{Conclusions and further work}

In this paper we have described replacing the existing P2P communications in the real world atmospheric model, MONC, with MPI RMA. Designed around the concepts of components, multiple MONC components perform halo-swapping by calling into a central module of the model core which abstracts them from the complex underlying mechanism of communication. For compatibility it was important that the halo-swapping API provided by the model core, designed with P2P in mind, was not modified. We replaced P2P with MPI RMA levering all three synchronisation modes; fences, PCSW and passive target synchronisation. We have described our approach for retrofitting MPI RMA, the considerations we had to bear in mind for correctness and modifications required for optimisation. 

We have considered performance and scalability on up to 32768 cores of ARCHER, a Cray XC30 and observed - crucially with DMAPP enabled, there is a benefit of replacing the existing P2P communications with MPI RMA.  The performance of PSCW and passive target synchronisation is comparable and far better than the other mode of RMA synchronisation, fences, which itself does not outperform MPI P2P. In contrast to the Cray MPI RMA implementation (with DMAPP enabled), the SGI MPI Toolkit (MPT) support for RMA is less mature and on Cirrus, an SGI ICE machine with Infiniband interconnect, MPI P2P outperformed all RMA configurations. We therefore conclude that MPI RMA is not a communication panacea, but PSCW and passive target synchronisation especially does exhibit some benefits in a real world application as long as library and hardware support is sufficiently mature, as is the case with Cray, and code modifications are done appropriately.

In terms of evaluating MPI RMA we have focused on the performance and scalability, but another consideration is the programmability. MPI RMA is less well known than P2P communications and arguably such concepts as multiple memory models can confuse programmers. For this work we found that the enhancements to MPI RMA in version 3.0 of the standard, ones that we especially rely on for our passive target synchronisation implementation, are welcome and improve support. We also found a number of excellent resources for learning MPI RMA at \cite{adv-mpicourse} and \cite{cornell-mpi}. In our opinion, once the programmer works through these resources and learns the basic concepts, MPI RMA is not much more complex than MPI P2P. But crucially P2P and RMA are very different and need to be approached with that understanding in mind. The programmer must embrace the fact that RMA is built upon different concepts which they should understand separately, rather than trying to understand MPI RMA in the context of their existing P2P knowledge.

In terms of future work, in this paper we have concentrated on MPI but there are other single sided technologies such as GASPI \cite{gaspi}. Research has shown that there are potential performance benefits to be gained by using GASPI rather than MPI \cite{gaspi-performance} and GASPI's \emph{notify} mechanism could be a very useful and optimal way of informing the target when data has been remotely written and hence can be used. A challenge with this would be the inevitable interoperability issues between MPI and GASPI that would need to be faced, as other parts of the MONC code rely on MPI (e.g. parallel IO \cite{in-situ}.) However work is being done on interoperability between these technologies \cite{gapsi-mpi-bpg} which might ease such an investigation. In terms of improvements to MPI RMA itself, we believe that a similar mechanism to GASPI which combines the movement of data with some form of notification would be very useful and help address the consistency issues discussed in this paper.

\appendices
\section*{Acknowledgements}

This work was funded under the embedded CSE programme of the ARCHER UK National Supercomputing Service (http://www.archer.ac.uk) This work used the ARCHER UK National Supercomputing Service (http://www.archer.ac.uk) and the Cirrus UK National Tier-2 HPC Service at EPCC (http://www.cirrus.ac.uk).

\end{document}